\documentclass[aps,prl,twocolumn,groupedaddress]{revtex4}
\usepackage{graphicx}
\begin{document}
\title{Structural flexibility of cuprate superconductors}
\author{Kostya Trachenko}
\email[]{kot@esc.cam.ac.uk}
\affiliation{Department of Earth Sciences, University of Cambridge,
Downing Street, Cambridge, CB2~3EQ, UK}

\begin{abstract}
We study the lowest energy state in the CuO$_2$ plane of cuprate
superconductors, related to the vibration that does not involve
distortions of constituent units of the plane, the Rigid Unit Mode
(RUM). We discuss the correlated motion in the plane due to RUM on
temperature decrease, and possible relevance of this phonon for
superconductivity.
\end{abstract}

\pacs{PACS numbers:}

\maketitle

The existence of the hierarchy of interatomic interactions in a
solid can make a substantial reduction in analyzing its properties.
If the energy to break a chemical bond between two atoms
considerably exceeds the thermal energy, such a bond can be viewed
as a Lagrangian constraint, in a sense that it keeps two atoms at a
fixed distance. This idea can be made useful in the case of covalent
materials, in which two-body stretching and three-body bending
forces considerably exceed all others. These short-range
interactions can be translated into the building blocks of a
mechanical network. This has been the starting point of the Phillips
theory of network glasses \cite{phillips}. By requiring that the
number of degrees of freedom is equal to the average number of
bonding constraints, this theory predicted the average coordination
number of $\langle r\rangle =2.4$ for which glass forming ability is
optimized. Since then, this picture has been widely used to discuss
relaxation in covalent glasses and crystals.

The constraint theory offers a great reduction in treating the
interactions in a solid, by translating static, vibrational and
relaxation properties of a solid into those of a mechanical network,
with well-developed methods to study it. For example, the procedure
known as Maxwell counting can make rigorous predictions about the
low-energy states of the system. Any modes in a mechanical network
that keep local constraints (e. g. two-body stretching and
three-body bending constraints) intact, have zero frequency because
there is no restoring forces to such deformations. According to
Maxwell counting, the number of such modes is equal to the
difference between the number of degrees of freedom, $N_f$, and the
number of bonding constraints, $N_c$. Therefore, the existence of
the hierarchy of interactions in a solid can have important
implications for the hierarchy of vibrational modes in terms of
their frequency. In the simulation study of constraint counting in
glasses, ``floppy'' modes appear when the network becomes
under-constrained, $N_c < N_f$, or $\langle r\rangle <2.4$ \cite{he}
(the term ``floppy'' here points to the fact that in real systems,
weaker interactions always give a non-zero restoring force
associated with propagation of constraint-obeying modes, making
their frequency not zero exactly, but some small values).

By construction, the picture which maps interatomic interactions
into a network of mechanical constraints, is limited to solids with
short-range covalent interactions. If ionic contribution to bonding
is substantial, the mapping of interatomic interactions into a
network is problematic due to the long-range nature of Coloumb
forces and the absence of the hierarchy of interactions
\cite{thor-coll}. It is nevertheless still possible to consider many
important properties of solids with substantial ionic contribution
to bonding, using a very general idea that a certain chemical
interaction can be mapped into a mechanical constraint. Consider
very common silica glass. Ionic contribution to Si-O bond is at
least as strong as covalent one, resulting in the fact that although
O-Si-O bending constraint is intact, Si-O-Si angular constraint is
broken, as is seen by the very broad distribution of Si-O-Si angles
\cite{mozzi}. Hence the usual constraint counting procedure would
overestimate rigidity of silica glass. However, despite the
substantial ionicity of Si-O bond, it is known from both experiments
and computer simulations that SiO$_4$ tetrahedra are very rigid.
This is related to the high energy cost involved in the deformation
of the electronic density that has a tetrahedral symmetry. Hence if
we are interested in low-energy vibrations of silica, its Phillips
network analogue is a collection of SiO$_4$ rigid units, loosely
connected at corners. Thus even though there is a considerable ionic
contribution to bonding in a solid, the knowledge of its structure
and chemistry can still allow us to map interatomic interactions
into a generalized network, albeit with different building blocks:
these do not correspond to two- and three-body constraints as in the
Phillips theory, but to local rigid units. The modes that propagate
in without its constituent units having to distort have been named
Rigid Unit Modes (RUMs) \cite{dove-rums}.

In this paper, we discuss the ability of cuprate high-temperature
superconductors to support the RUM, which we associate with the
lowest non-trivial energy state in CuO$_2$ plane. A distinct
property of the RUM is its infinite (in the idealized model)
correlation length. We discuss the behaviour of the RUM on
temperature decrease, and its possible relevance for
superconductivity.

We consider the common structural unit of cuprate superconductors,
the CuO$_2$ plane. In order to discuss the ability of this system to
support RUMs, one needs to identify rigid units, analogous to
SiO$_4$ tetrahedra in silica. Cuprate superconductors are materials
with mixed covalent and ionic bonding, hence, unlike in silica,
their structures do not immediately offer the way to map them into
the collection of rigid units, which reflects the bonding type. A
useful insight comes from the experimental and quantum-mechanical
results that there is substantial covalency in Cu--O bonding in the
Cu--O plane \cite{tajima}. Next, as in silicates, experiments point
to the broken O bond-bending constraints \cite{phil2}. This allows
us to consider the two-dimensional system of corner-shared rigid
CuO$_4$ squares, loosely connected at corners (compare to a
silicate, modeled in the RUM model as the system of rigid SiO$_4$
units, loosely connected at corners).

For this system, Maxwell counting gives the result that it is
over-constrained, $N_c>N_f$. Indeed, each square has three degrees
of freedom in the plane, two translational and one rotational,
giving $N_f=3$. There are two constraints per shared corner, or one
constraint per corner per CuO$_4$ unit, or 4 constraints per unit in
total, $N_c=4$. Hence Maxwell counting predicts that no RUM-type
distortions should exist in CuO$_2$ plane. However, this approach
does not take into account the important property of the system, its
symmetry. It has been realized that symmetry can make certain
constraints redundant, resulting in the effective increase of system
flexibility against RUMs. This can be studied in some detail using
the lattice dynamics approach to analyze the flexibility of a system
of connected rigid units, loosely connected at corners. This method
rigorously finds all RUMs in a given system, in addition to three
trivial acoustic modes at $k=0$ \cite{dove-rums}. For the perovskite
system, it has been shown that symmetry reduces the number of
constraints in such a way as to give one particular RUM that
corresponds to rotations and displacements of rigid units
\cite{dove-trac}.

We can readily extrapolate this result to the two-dimensional
analogue of the perovskite structure, the system of connected
CuO$_4$ squares. The corresponding mode, the optic RUM, is shown in
Figure 1.

\begin{figure}
\begin{center}
{\scalebox{0.75}{\includegraphics{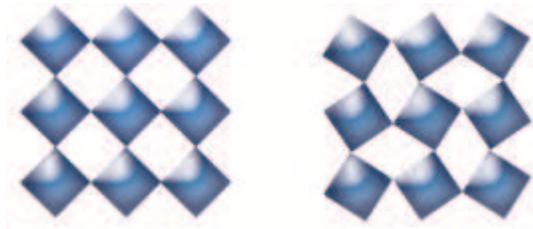}}}
\end{center}
\caption{The Rigid Unit Mode (RUM) in the CuO$_2$ plane of cuprate
superconductors.} \label{fig1}
\end{figure}

It is interesting to note that negative thermal expansion, decrease
of volume (or some of the system's linear dimensions) on temperature
increase, is a clear signature of the presence of RUMs in a system.
Indeed, distortion shown in Figure 1 pulls the structure onto
itself, and if this effect exceeds thermal expansion, the net effect
can be volume reduction on temperature increase. Using this picture,
its has been possible to explain negative thermal expansion in a
variety of RUM-floppy materials \cite{nte}. This effect is also seen
in the in several cuprates as anomalous change of the unit cell
parameters in CuO$_2$ plane. For example, in
YBa$_2$Cu$_3$O$_{7-\delta}$, deviation from the linear decrease of
the parameters of CuO$_2$ plane is seen around superconducting
temperature $T_c$, followed by the negative thermal expansion at
lower temperature \cite{youprb}. We can therefore interpret these
effects as the manifestation of existence of the RUM in the CuO$_2$
plane.

That cuprate superconductors can support the mode shown in Figure 1
(referred to as the RUM in the discussion below) is not a new
result. What is important, however, is the method of its derivation,
from developing the idea about the hierarchy of interactions in the
solid into the algorithm that rigorously finds all RUMs in a given
system. By this construction, the RUM is the lowest energy
vibrational state in the CuO$_2$ plane (apart from trivial acoustic
modes), because it does not involve bond-stretching and bond-bending
distortions that come at high energy cost. Any other non-trivial
in-plane mode has necessarily higher frequency because it involves
bending and stretching bonds (for example, the energy of the
bond-stretching breathing mode is 18--20 THz \cite{pint-rev},
several times larger than the RUM frequency, see below). This has
implications for the low-energy correlated motion in the CuO$_2$
plane, as we discuss below.

An important feature of the RUM shown in Figure 1 concerns its
correlation length $\zeta$, which enters the coordinate correlation
function $\left< q({\bf r})q(0) \right> \propto \exp(-r/\zeta)$. In
the model in which CuO$_4$ squares are rigid, $\zeta=\infty$,
reflecting the fact that rotation of one square necessarily triggers
the motion (rotations and displacements) of squares in the whole
plane (see Figure 1). To appreciate this property, it is instructive
to consider two different types of systems for which this is not the
case. An example of the first type is the system in which triangles,
as opposed to squares, are loosely connected at corners to form the
network of six-member rings. The RUM model, applied to this system,
gives the result that unlike the system of connected squares, there
are many of ways in which such a system can distort by local {\it
independent} RUMs. In other words, the connected system of triangles
is much more RUM-floppy than that of the squares. This can be
understood using the Maxwell counting discussed above: the number of
degrees of freedom for the system of triangles is the same as for
the square system, $N_f$=3 per unit, but the number of constraints
per unit, $N_c=(3\cdot2)/2=3$, is smaller than that for the square
system ($N_c$=4). As a result, the correlation length is limited to
several unit cells only: it is easy to show that for the system of
connected triangles, rotation of one triangle involves rotations and
displacements of triangles within the radius of 2-3 ring diameters
only. Another example of a system of the first type is the
three-dimensional random network of tetrahedra, loosely connected at
corners. Similar to the previous example, the RUM model gives the
result that there are many ways in which this system can distort by
independent RUMs \cite{prl}. As a result, the correlation length
$\zeta$ is only about 10 \AA, as deduced from the neutron scattering
experiments \cite{bermejo}. The second type of systems are very
different from the first type: they do not support RUM at all. The
structure of silicon is one such example. Here, unlike in SiO$_2$,
all bending constraints are intact, and the structure is severely
over-coordinated to allow for existence of any RUMs.

From the topological point of view, one can therefore view the
two-dimensional system of connected squares in Figure 1 as an
interesting borderline and, perhaps, unique, case for its ability to
support low-energy vibrations. It is neither under-constrained as to
give many RUMs and hence small correlation length, nor
over-constrained as to inhibit RUMs altogether. The balance of the
degrees of freedom and the number of constraints, together with the
system's symmetry, give the result that only one single RUM is
present (see Figure 1), which has the infinite correlation length.
It should be noted that $\zeta=\infty$ is true only in the model in
which the CuO$_4$ units are infinitely rigid. In practice, there is
always a finite distortion of the units, leading to a finite $\zeta$
which, however, considerably exceeds the size of the unit. In this
case, on temperature decrease (see below), the correlated atomic
motion in the plane takes place in the domains of size $\zeta$.
Generally, $\zeta$ can be affected by the multitude of interactions
present in the cuprate system at low temperature.

The frequency of the RUM mode is zero only in the RUM model, and in
real cuprates, $\nu_{\rm r}$ is defined by the inter-unit and other
next-order interactions, as well as by the effects of steric
hindrance on the CuO$_2$ plane. During the rhombic distortion, one
pair of O atoms comes closer to the out-of-plane cations directly
above the rhomb centre (for example, in La$_2$CuO$_4$, these are
La/Sr ions; in YBa$_2$Cu$_3$O$_7$, these are Y and Ba ions), whereas
the remaining pair comes closer to the cations above the centres of
neighbouring rhombs. The energy of these interactions sets the scale
of $\nu_{\rm r}$.

A large body of data exists on phonons in cuprates. Most of the
recent discussion has concentrated on the higher-energy breathing
and half-breathing modes, whereas the RUM is not commonly discussed.
A possible reason for this is that this low-frequency mode is hardly
visible in the total density of states \cite{ourjop}. In earlier
studies \cite{pint-rev,reich}, the combination of neutron scattering
experiments with lattice dynamics calculations has identified the
RUM as the zone boundary optic phonon with $\nu_{\rm r}$ in the
3--3.6 THz range in the tetragonal phase of Nd$_2$CuO$_4$,
Pr$_2$CuO$_4$, La$_2$CuO$_4$, and 6.3 THz in YBCO.

We can now discuss the vibrational behaviour in the CuO$_2$ plane as
temperature decreases. We first note that the values of $\nu_{\rm
r}$ above are considerably higher than in other RUM-floppy systems
(for example, in silicates, $\nu_{\rm r}$ is in the range 0--1 THz
\cite{ourjop}). One can therefore expect that temperature $T_{\rm
r}$, at which transition of the RUM to the ground state takes place,
is relatively high. $T_{\rm r}$ can be calculated from the
order-of-magnitude estimation that the temperature-dependent term in
the RUM energy, $\langle E_{\rm r}\rangle=\frac{1}{2}h\nu_{\rm
r}+h\nu_{\rm r}/(\exp(h\nu_{\rm r}/kT)-1)$, is ten times smaller
than the ground state energy, giving $\exp(h\nu_{\rm r}/kT)-1=20$
and $kT_{\rm r}=0.33h\nu_{\rm r}$. Taking $\nu_{\rm r}$ from the
3--6.3 THz range above, we obtain $T_{\rm r}$ in the 48--101 K
range, and we remark here that $T_{\rm r}$ is on the scale of $T_c$
of cuprate superconductors. An approximate estimate for the
vibrational amplitude $q$ in the RUM ground state gives a relatively
large value, $q=(h/2m\nu_{\rm r})^{1/2}$=0.18--0.26 \AA, if $m$ is
the mass of the CuO$_4$ unit.

At high temperature $T>T_{\rm r}$, atomic motion in the plane is the
superimposition of all phonons in their excited states. Although the
RUM is present, the atomic displacements are de-correlated by other
high-energy phonons with small correlation lengths. By construction,
the RUM is the lowest energy state in the CuO$_2$ plane, apart from
the acoustic modes. Therefore at $T_{\rm r}$, all the phonons in the
higher-frequency optic branches freeze in their ground states. These
phonons, with the upper frequency of about 22 THz, constitute the
majority of vibrations in the cuprate density of states
\cite{pint-rev}. At $T_{\rm r}$, de-correlation effects due to these
phonons are decreased due to the reduced vibrational amplitudes in
their ground states, which, for higher-frequency phonons, are
further reduced as $\left<q^2 \right>\propto 1/\nu$. As a result,
the correlated motion on lengthscale $\zeta$, set by the RUM
(dominant in terms of the vibrational amplitude), increases. In this
respect, it is interesting to recall the ion channeling experiments
in several cuprates, which showed that the atomic motion becomes
highly correlated at the superconducting transition \cite{sharma}.
It should be noted that at $T_{\rm r}$, low-frequency acoustic
phonons remain to be in their excited states; however, as we discuss
below, de-correlation effects at long wavelengths may not be
important.

We now discuss possible relationship between the RUM, $T_{\rm r}$
and superconductivity. First, coupling between charge carriers and
the RUM is expected to be large. In BSCCO, for example, the spectral
function shows strong coupling to modes in the 12--27 meV range
\cite{gonnelli} ($\approx$3--6 THz), which is close to the range of
$\nu_{\rm r}$. Second, one can tentatively discuss how the RUM is
related to the phase coherence between the charge carriers (e.g.,
Cooper pairs). Suppose a pair is formed as the result of an electron
(hole) scattering the RUM phonon at, for example, O site, and
another electron absorbing it at a distance $\xi$. If, at the same
time, another pair is formed in the same way in a different region
of the plane, but within the distance $\zeta$ from the first pair,
the two pairing events are identical due to the correlated RUM
motion (see Figure 1), resulting in the correlation between the
phases of the pairs. In this sense, the RUM has no bearing on the
coherence of the Cooper pairs if $\xi\ll\lambda$, as in the case of
$\xi$ for cuprates and $\lambda$ for long-wavelength acoustic
phonons. An interesting observation is that in cuprates, the size of
the Cooper pair, $\xi$ (13 \AA\ in YBCO \cite{mourach}), is
comparable with the RUM modulation length, $\lambda_{\rm r}\approx
2\sqrt2d\approx 11$ \AA, where $d$ is the O-O distance (see Figure
1).

The relationship between the RUM, $T_{\rm r}$ and superconductivity
discussed above needs further study. However, if this relationship
exists, it is consistent with several broad experimental
correlations. First, one expects that $T_c$ should be maximized in
the system in which the RUM is favoured most. Experimentally, for
different cuprates at a given doping level, $T_c$ is maximal in
structures in which the CuO$_2$ plane is flat and square
\cite{mourach}. In our picture, this behaviour can be rationalized
by noting that the RUM is most favoured when the CuO$_2$ plane is as
close to being flat and square as possible. On the other hand,
distortion of the plane results in suppression of the RUM (see
Figure 1). The distortion can be plane buckling due to either static
or dynamic tilting motion. In this case, suppression of the RUM is
caused by the out-of-plane strains, although these can be partially
compensated by slightly different Cu--O distances \cite{bianconi}.
Another type of distortion can be the rhombic distortion of the flat
plane, which also suppresses the RUM due to the lower symmetry of
the orthorombic phase relative to the tetragonal one. Second, one
expects that $T_c$ should increase with $\nu_{\rm r}$, since a
larger $\nu_{\rm r}$ implies a higher temperature of the onset of
correlated motion, $T_{\rm r}\propto \nu_{\rm r}$. Experimentally,
cuprates show unconventional relationship between $T_c$ and lattice
stiffness: unlike in strongly coupled BCS superconductors, in
cuprates, $T_c$ increases with lattice stiffness, or Debye
temperature, within a given cuprate group \cite{mourach}. In our
picture, this behaviour can be rationalized by noting that $\nu_{\rm
r}$ (and hence $T_{\rm r}$) increases with lattice stiffness. Third,
pressure, stiffening the lattice, increases $\nu_{\rm r}$. In our
picture, this should result in the increase of $T_c$, and is
consistent with experimental data. Fourth, one expects to find a
positive isotope shift due to atoms involved in the RUM motion, i.e.
in-plane atoms. This is consistent with experiments that show that
the predominant contribution to the (positive) isotope shift comes
from the planar atoms \cite{zhao}. It is also found that the isotope
shift decreases with $T_c$ \cite{zhao}. In this respect, it is
interesting to note that a smaller atom mass, in addition to
increasing $T_c$ (giving the positive isotope shift $T_c\propto
1/m^\alpha$), also increases $\left<q^2 \right>\propto 1/m\nu\propto
1/\sqrt m$, therefore increasing de-correlation effects of
higher-energy phonons. In our picture, this decreases $T_c$, and
this effect is expected to be more pronounced at higher
temperatures.

Finally, it is interesting to note that in cuprates, the charge in
the CuO$_2$ plane is dynamically ordered in stripes that run either
along Cu-Cu-Cu or Cu-O-Cu bonds \cite{mourach}. This pattern is
consistent with the field of the RUM displacements (see Figure 1).
This hints to a possible relationship between the RUM-induced field
and the dynamic stripe order. This picture may be the subject of
further study that also includes other interactions (e.g., magnetic)
present in the system.

In summary, we discussed the property of cuprate superconductors to
support modes that do not involve distortions of constituent units.
The corresponding mode, the RUM, is the lowest non-trivial energy
state in the CuO$_2$ plane. We discussed the correlated motion in
the plane due to RUM on temperature decrease, and possible relevance
of this phonon for superconductivity.

I am grateful to Prof. M. T. Dove, E. Artacho, W. Lee, J. C.
Phillips and V. V. Brazhkin for discussions, and to EPSRC for
support.

\end{document}